\documentclass[sigconf, authorversion]{acmart}
\usepackage{todonotes}

\title[From Automation to Cognition]{From Automation to Cognition: Redefining the Roles of Educators and Generative AI in Computing Education}

\author{Tony Haoran Feng}
\orcid{0009-0005-8608-6961}
\affiliation{
  \institution{University of Auckland}
  \city{Auckland}
  \country{New Zealand}
}
\email{hfen962@aucklanduni.ac.nz}

\author{Andrew Luxton-Reilly}
\orcid{0000-0001-8269-2909}
\affiliation{%
  \institution{University of Auckland}
  \streetaddress{Private Bag 92019}
  \city{Auckland}
  \country{New Zealand}
}
\email{a.luxton-reilly@auckland.ac.nz}

\author{Burkhard C. W\"{u}nsche}
\orcid{0000-0002-8013-4118}
\affiliation{%
  \institution{University of Auckland}
  \streetaddress{Private Bag 92019}
  \city{Auckland}
  \country{New Zealand}
 }
\email{burkhard@cs.auckland.ac.nz}

\author{Paul Denny}
\orcid{0000-0002-5150-9806}
\affiliation{%
  \institution{University of Auckland}
  \streetaddress{Private Bag 92019}
  \city{Auckland}
  \country{New Zealand}
 }
\email{paul@cs.auckland.ac.nz}

\copyrightyear{2025}
\acmYear{2025}
\setcopyright{rightsretained}
\acmConference[Australasian Computing Education Conference]{Submitted to ACE 2025}{February 2025}{Brisbane, Australia}
\acmBooktitle{ACE '25}
\acmDOI{}

\begin{document}

\begin{abstract}


Generative Artificial Intelligence (GenAI) offers numerous opportunities to revolutionise teaching and learning in Computing Education (CE). However, educators have expressed concerns that students may over-rely on GenAI and use these tools to generate solutions without engaging in the learning process. While substantial research has explored GenAI use in CE, and many Computer Science (CS) educators have expressed their opinions and suggestions on the subject, there remains little consensus on implementing curricula and assessment changes.

In this paper, we describe our experiences with using GenAI in CS-focused educational settings and the changes we have implemented accordingly in our teaching in recent years since the popularisation of GenAI. From our experiences, we propose two primary actions for the CE community: 1) redesign take-home assignments to incorporate GenAI use and assess students on their process of using GenAI to solve a task rather than simply on the final product; 2) redefine the role of educators to emphasise metacognitive aspects of learning, such as critical thinking and self-evaluation. This paper presents and discusses these stances and outlines several practical methods to implement these strategies in CS classrooms. Then, we advocate for more research addressing the concrete impacts of GenAI on CE, especially those evaluating the validity and effectiveness of new teaching practices.
\end{abstract}



\maketitle

\section{Introduction}
Generative Artificial Intelligence (GenAI) has profoundly impacted the Computing Education (CE) landscape \cite{denny2024cacm}. Studies have shown that GenAI is capable of producing solutions for introductory and intermediate (CS1/CS2) programming assessments~\cite{denny2023conversing, finnie2022robots, savelka2023can}, with recent models achieving near-perfect scores on CS1 exams \cite{prather2023robots}. Furthermore, OpenAI's GPT-4o model has demonstrated the potential to mimic human educators and provide tutoring independently~\cite{gpt4o_khan}. Leveraging both capabilities may provide Computer Science (CS) students with unlimited, immediate, and personalised tutor assistance, which can also reduce the workload of answering student queries for CS educators~\cite{kazemitabaar2024codeaid, sheese2024patterns}. GenAI may also be particularly useful for subjects using multimodal input and output. For example, in Computer Graphics, the correctness of a student submission can be assessed using visual output, function output, API invariants (OpenGL states), and code \cite{wunsche2018automatic, wunsche2019coderunnergl}. In the future, GenAI tools may be able to give formative feedback on all of these levels. This could help students develop a deeper understanding of content, and enable educators to pose more open-ended and creative assignments, which would be difficult to automatically assess with more traditional means~\cite{hooper2024advancing}.

While the capabilities and rapid adoption of GenAI are impressive, it is a double-edged sword for CS students and educators. GenAI tools accept an unlimited number of queries and produce immediate responses. This permits students to repeatedly seek assistance until they receive working solutions, even if they don't understand the underlying concepts, potentially becoming overly reliant on GenAI to generate answers without actively engaging in the learning process~\cite{becker2023programming, prather2023robots}. Additionally, despite GenAI's impressive ability to solve a wide range of problems, it is currently less effective when applied to certain complex questions, such as those in Computer Graphics~\cite{feng2024more, feng2024eye, feng2024can}, where students may learn from incorrectly generated information if they use GenAI to aid them in their problem-solving process. Unfettered access to tools that can be used by students to solve exercises with minimal effort threatens the integrity and effectiveness of take-home assessments.
To avoid this, some educators discourage using GenAI on graded work and attempt to detect and punish any misuse of GenAI tools~\cite{tang2024science, wu2023survey}. However, GenAI detection tools are unreliable and could result in unjust punishments~\cite{orenstrakh2023detecting}. Alternatively, educators could adapt course content and teaching practices that encourage productive learning with unrestricted use of GenAI in assignments~\cite{milano2023large}, balanced by the use of secure invigilated assessments to ensure valid credentialing of learning~\cite{lye2024generative}.

As GenAI continues to impact the CE landscape, action must be taken to maintain a productive, fair, and effective learning environment for CS. Recent research involving comprehensive interviews with CS educators worldwide found that a common strategy to prevent student misuse of GenAI is to rely more on in-person invigilated assessments and less on take-home assignments while shifting to grading criteria that assess processes instead of the final product~\cite{prather2023robots}, a sentiment that is widely shared~\cite{cao2023navigating}. These articles suggest that educators should teach skills that GenAI cannot teach, such as replicating human judgment. However, concrete methods to implement these strategies must still be developed, evaluated and deployed at scale.

In this paper, we recount some personal anecdotes when navigating CE after the popularisation of GenAI while highlighting several issues brought by GenAI to CE (Section \ref{early}). We propose two suggestions to the CE community to address these issues, along with justifications and the recommended courses of action to implement these suggestions in CS classrooms (Section \ref{discussion}).

\section{Early Experiences}
\label{early}
This section describes our experiences using GenAI in CS-focused educational settings and the changes we have implemented accordingly in our teaching in recent years since the popularisation of GenAI.

\subsection{Context}
We teach at a large urban research-focused institution.  Authors are primarily involved in teaching classes of approximately 1000 first-year Engineering students learning to program in C, and classes of approximately 300-500 first-year Computer Science students taught in Python.  In both cases, students are engaged in weekly laboratory activities conducted in person and supported by graduate teaching assistants.  The laboratory activities are submitted to \textit{CodeRunner}, where the correctness of programming solutions is graded automatically using a test suite.  Students also engage in other activities manually graded by the teaching assistants or involving other online tools (e.g., those described later in this paper).  At the undergraduate level, authors also have experience teaching second-year systems, and third-year computer graphics courses.

\subsection{Student Use of GenAI}
Since the release of ChatGPT~\cite{chatgpt}, we have seen a steady increase in student use of GenAI in CS classrooms. During the first few months after ChatGPT's release, students were still wary of the adverse consequences of using GenAI in classrooms, as the majority of students were still uncertain of its performance and accuracy or were unaware of its existence, and those who knew its capabilities were concerned with potential violations of academic integrity through its use.

However, as GenAI grew in popularity, more students became aware of its capabilities, and the use of ChatGPT and other GenAI-powered tools became more prevalent. Despite the guidelines that prohibited using GenAI on course pages and assignment/assessment specifications, there was evidence that many students still used GenAI against the guidelines. For example, the average scores achieved by students for take-home assignments were much higher than those for in-person invigilated assessments. Furthermore, we observed in assessments of an introductory programming course that a few students struggled to write simple print statements despite having completed all prior take-home programming assignments, often using sophisticated code constructs and solution approaches.  There were also a small number of cases of students accessing tools like ChatGPT during invigilated assessments such as tests and quizzes, in violation of the rules.  We are aware that there are many reasons that students struggle and the difficulties observed are not all necessarily caused by GenAI. Even so, publicly available tools, such as ChatGPT, have made it possible to complete some take-home assignments without engaging in the learning process by providing answers to coursework at no cost or risk. This enables less motivated students to generate solutions and submit them as their own without engaging in learning. However, in contrast, we found that more motivated students were less likely to misuse GenAI and instead use it to their advantage to support their learning, such as by asking constructive questions and receiving immediate feedback to clarify confusion or misunderstanding that may arise during learning.  Our observation aligns with the work of Prather et al., who found that while tools like ChatGPT and Copilot can help motivated students accelerate their work on programming tasks, struggling students often face persistent and compounded metacognitive difficulties when using these tools \cite{prather2024widening}.  As a result of our own observations, we saw significant importance in teaching students effective ways of using GenAI to support their learning. 

As time passed and GenAI became ubiquitous, it became clear that policy changes had to be made to adapt, and simply prohibiting the use of GenAI was not the solution. In fact, the prohibition of GenAI during an academic programme may cause difficulty for graduates expected to transition to work environments where GenAI may be encouraged to boost productivity. Therefore, we began to allow, and even incorporate, GenAI in take-home assignments (several concrete examples are given in Subsection~\ref{prompt}), so it became commonplace for students to use GenAI, even in the presence of teaching staff. But to ensure that students were engaging in learning and not reliant on GenAI, we enforced strict standards for some assessments, particularly secure (in-person, invigilated, and GenAI-prohibited) or online using secure assessment tools. Furthermore, we highlighted to students the nature of the assessments and, thus, the importance of learning and understanding the concepts even with the use of GenAI.

\label{queries}
Anecdotally, over time  we have noticed a reduction in student queries regarding concepts or questions related to programming exercises, and relatively higher percentages of student queries related to course administration. This is presumably because many students are making use of GenAI for receiving fast and often detailed answers.



\begin{figure}[t]
    \centering
    \includegraphics[width=\linewidth]{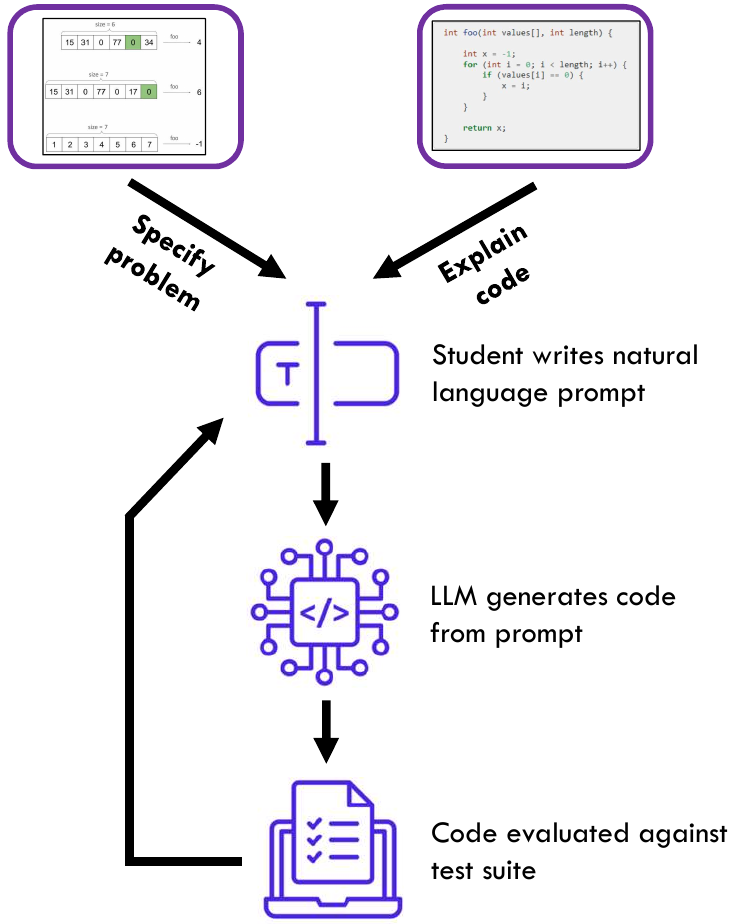}
    \caption{Providing automated feedback to students on their natural language prompts using code generation from an LLM.  This pipeline has been applied to solving computational tasks (``specify problem'') where the student is shown a visual depiction of a problem, and to code comprehension tasks (``explain code'') where the student is shown a code fragment to explain.}
    \label{fig:schematic}
\end{figure}

\subsection{Incorporating GenAI into the classroom}
\label{prompt}

As students become increasingly familiar with using commercial AI tools, like ChatGPT, we recognise the importance of integrating GenAI into classroom activities to align with their expectations.  Over the last few years, we have explored several  activities that leverage large language models (LLMs) to provide unique and engaging ways for students to interact with computing concepts.  These activities aim to develop students' problem-solving, debugging, and conceptual understanding skills through meaningful interaction with LLMs. To develop critical thinking skills, we have also explored scaffolded tasks where students seek help from LLM-based teaching assistants and evaluate their outputs.  Below, we outline six key activities explored in our first-year introductory programming courses.

\subsubsection{Prompt Problems}
\label{promptproblems}
Prompt Problems~\cite{denny2024prompt} are a relatively new kind of  exercise designed to teach students the emerging skill of writing effective prompts for generating code. These exercises require students to craft natural language prompts that enable an LLM to generate correct code for solving a specified computational task. The activity emphasizes computational thinking and critical evaluation of AI-generated outputs rather than direct programming skills. 

In our deployment of Prompt Problems, students were presented with visual representations of programming tasks, encouraging them to infer and articulate requirements in a prompt.  This is depicted in the ``Specify problem'' branch of Figure \ref{fig:schematic}, where the example image shows a series of `input' arrays along with their expected `outputs'.  We used a web-based tool called \textit{Promptly} to support these exercises by automating the evaluation of generated code against test cases. Results from our implementation showed that students were highly engaged, finding the activity challenging and rewarding.  Students liked that the problems introduced them to new programming constructs and techniques.  Importantly, these tasks allow students to refine their problem-specification skills, a critical aspect of AI-assisted programming.

\subsubsection{Generating Debugging Exercises}
Debugging is a critical skill for programmers, and we have used the BugSpotter~\cite{padurean2024bugspotter} tool  as an innovative way to teach this skill. The tool generates buggy code based on a problem specification and requires students to design failing test cases that reveal the errors. This unique format helps students develop systematic reasoning about code and reinforces the importance of understanding problem specifications.

We deployed BugSpotter in one of our large introductory programming courses, and conducted an evaluation of student performance solving the tasks and expert review of the generated problems. The LLM-generated debugging exercises were comparable in quality to instructor-designed tasks, demonstrating that AI can effectively support the creation of diverse sets of exercises with varying difficulty.  There are other emerging LLM-based approaches for teaching debugging, including having students play the role of a teaching assistant to help an LLM-powered agent debug code~\cite{ma2024hypocompass}. Similar to our approach, this encourages metacognitive practices, as students are required to hypothesize the cause of errors and then verify their solutions. This also aligns with the growing need to prepare students for debugging AI-generated code in real-world scenarios.


\subsubsection{Code Comprehension through Prompting}
Code comprehension is an essential skill for novice programmers, especially in an era where the volume of AI-generated code is rapidly increasing.  We have been using Explain in Plain Language (EiPL) tasks~\cite{kerslake2024integrating} to help students practice reading and understanding code by receiving automated feedback from an LLM. These tasks require students to explain the high-level purpose of a provided code fragment, by writing a prompt that will generate code equivalent to the original.  This dual focus -- understanding code and articulating problems in natural language -- bridges the gap between problem-solving and programming.

Our implementation is depicted in the ``Explain code'' branch of Figure \ref{fig:schematic}, where the exercise begins with the student being shown a fragment of code (typically a single function with obfuscated identifier names).  In the course, we interleaved EiPL tasks with traditional programming-based lab activities. Students who struggled with syntax in traditional tasks found the natural language-focused EiPL exercises more accessible. Performance on these tasks was less correlated with traditional programming skills, indicating that they target a complementary skill set. Additionally, students appreciated the immediate feedback provided by the LLM on their prompts, which facilitated iterative learning. 

\subsubsection{Student-Generated Analogies}
Understanding complex concepts such as recursion often poses challenges for students. To address this, we implemented a novel exercise where students used LLMs to generate analogies to help them understand recursion ~\cite{bernstein2024nesting}.  In order to produce useful analogies, students were encouraged to use themes that aligned with their personal interests.  We also asked students to critically evaluate the analogies that were generated, and reflect on the usefulness of this approach.  Our goal in having students produce analogies was to help them anchor abstract concepts in familiar, personally relevant contexts, making them more accessible.

In a study involving over 350 students, participants used ChatGPT to create analogies for recursion. Results showed that analogies tailored to students' interests were more diverse and memorable than generic examples. Students reported that the activity deepened their understanding of recursion and made the learning process more enjoyable, with some students indicating they planned to use analogy generation in the future. This exercise demonstrates the potential of LLMs as tools for scaffolding personalized learning experiences.

\subsubsection{Generating Contextualized Parsons Problems}
Parsons Problems, which involve rearranging code fragments into a correct sequence, are widely used to scaffold code-writing skills. Using a custom tool called PuzzleMakerPy~\cite{gutierrez2024automating}, we allowed students to generate personalized Parsons Problems. Students could select contexts and programming topics of interest, making the exercises more engaging and relevant.

The deployment of PuzzleMakerPy showed several benefits. Students reported high satisfaction levels with the personalized problems, noting that the contextual framing made the exercises more relatable.  The ability to select programming concepts was also reported as being useful for learning. From an instructional perspective, the tool significantly reduced the workload of creating diverse, high-quality problems. This work is one example in a growing body of literature in computing education around using LLMs to produce learning resources \cite{sarsa2022automatic, hou2024tailor, lekshminarayanan2024explainingcodeexamplesintroductory, macneil2023experiences, jury2024evaluating, 10342898}.

\subsubsection{Teaching Students to Use AI Teaching Assistants}
\begin{figure}[t]
    \centering
    \includegraphics[width=\linewidth]{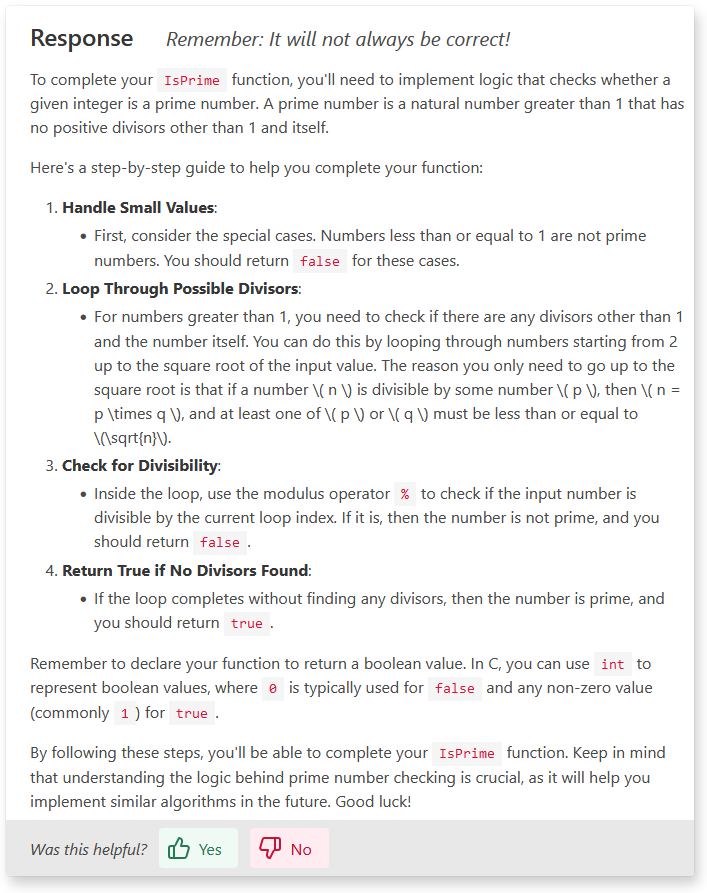}
    \caption{Example of a response from the CodeHelp digital teaching assistant to the query: \emph{``I want to write a function called IsPrime that returns true if the input value is a prime number.  Can you write the code for me?''}.  CodeHelp is designed to guide students with high-level suggestions rather than revealing direct code solutions.}
    \label{fig:codehelp}
\end{figure}

The advent of AI-powered teaching assistants offers scalable solutions for providing personalized feedback. In our courses we have deployed a digital assistant based on CodeHelp~\cite{denny2024desirable} to support students in debugging and code-writing tasks. The most obvious advantage of the AI assistant is that it could provide round-the-clock availability and ensure equitable access to support.  In fact, we observed students seeking help from the assistant at all hours, with hundreds of queries submitted between 1am and 3am during the two week period that the assistant was available.  The CodeHelp tool uses guardrails to avoid revealing direct code solutions -- an example of a response from the CodeHelp tool is shown in Figure \ref{fig:codehelp}.

Student feedback highlighted several key attributes of the AI assistant that were valued, including instant feedback, and scaffolding that guided students to discover their own solutions rather than simply providing direct answers. Students also expressed a desire for the responses from the AI assistant to be aligned with their current level of knowledge, which is a key skill that human tutors provide.  In general, our observations suggest that AI teaching assistants can complement human instructors by addressing gaps in availability and consistency. 

\subsubsection{Summary}
Together, these examples showcase the diverse applications of LLMs in computing education.  Integrating GenAI into classroom activities and into the curriculum is essential for equipping students with the skills to leverage AI tools in the future.

\section{Discussion}
\label{discussion}

We see two main approaches to incorporating GenAI into CS Education.  The first is to embrace GenAI as a core tool (similar to an IDE) used throughout a course.  An example of such a course redesign is described by \citet{vadaparty2024cs1-llm}, where the entire course has been refocused on developing programs with the integrated support of GenAI.  We believe this approach focuses on contemporary skills for workplace readiness.  The second approach is more moderate and retains the existing course design but adopts isolated activities introducing GenAI.  This approach focuses on raising student awareness of how GenAI may be used for learning.  

From our experiences navigating the introduction, popularisation, and acclimatisation of GenAI in CS education, we summarise two strategies that (a) introduce GenAI for individual tasks and (b) raise student awareness of how to use GenAI successfully for improved educational outcomes.

\subsection{Strategy 1: Redesign Assignments}
GenAI use is still not widely accepted in education, and using GenAI tools to assist with writing assignments is typically disallowed~\cite{perkins2023academic}. This is a sensible approach where the assignments have not been adapted for a GenAI context --- most assignments currently used in courses are designed to be solved individually and not with the assistance of GenAI. However, since GenAI tools are becoming more widely used, we believe that restricting GenAI use in all assignments is unrealistic and does not reflect real-world conditions. Hence, our view is that at least some assignments should be designed to support learning with GenAI. 

For example, the Prompt Problem exercises described in Subsection~\ref{promptproblems} require students to engage with problem specifications, an authentic task required by programmers writing programs to solve problems outside the highly constrained educational context. 
 This helps students develop important dispositions outlined in Computing Curriculum 2023~\cite{kumar2024computer}, such as being meticulous and persistent while raising awareness of the capabilities and limitations of GenAI tools.


Activities such as peer code review have been shown to benefit students producing reviews~\cite{indriasari2020review}.  These activities require students to understand and critique the code produced by others.  One potential barrier to such activities is the difficulty of sharing peer code in classrooms --- an administrative burden that typically requires a dedicated peer review tool.  The same activity (and indeed any activity that requires students to engage with work produced by others) can use GenAI to create the code to be reviewed.  Students can be asked to reflect on the quality of the code and comment on differences between the generated code and their solutions.  This approach retains the focus on traditional learning outcomes for a given assessment, yet it raises student awareness of GenAI capability (and potentially the limitations of such tools where the quality of the generated code has room for improvement).

As a general example, any traditional assignment involving code writing could be redesigned to focus on code comprehension and encourage students to link concrete examples more explicitly to more abstract concepts.  
Figure~\ref{fig:cs1-question} is an example of a typical code writing question used in a CS1 course. Some students may query GenAI for the solution and submit it as their work without processing the information. 
This can occur whenever we focus assessment on the \textit{product}, since GenAI is excellent at solving programming problems.  It is possible to redesign this question to focus on more abstract levels of understanding programming concepts and encourage students to use GenAI tools for scaffolding rather then simply producing solutions.  Figure~\ref{fig:cs1-redesigned} guides students to use the capabilities of GenAI tools as a scaffolding tool to support learning rather than replace learning. Although this does not guarantee productive learning, it lowers the likelihood of students copying and pasting without reading and understanding the responses. Assignments like this also align with the common suggestion of shifting towards assessing students on the learning and completion process rather than the final product~\cite{cao2023navigating, prather2023robots}.

\begin{figure}
    \centering
    \begin{tabular}{|p{0.45\textwidth}|}
        \hline
        Write a function that returns a function that takes an argument $x$ and adds it to the last even element in the given list. \\
        \hline
    \end{tabular}
    \caption{A typical programming question used in assignments for a CS1 course}
    \label{fig:cs1-question}
\end{figure}

\begin{figure}
    \centering
    \begin{tabular}{|p{0.45\textwidth}|}
        \hline
        Use the following guide to solve the following question: \\
        \textit{Write a function that returns a function that takes an argument $x$ and adds it to the last even element in the given list.} \\
        1. Use ChatGPT to find which programming concepts are used in this question. What are they? \\
        2. Ask ChatGPT to give an example of using each concept in a new context (unrelated to the problem above).  Summarise its responses in your own words. \\
        3. Write your solution to the question and then ask ChatGPT to generate a solution. \\
        4.  Compare the solutions and comment on the similarities and differences. Submit both solutions and your critique of the main differences between them.\\
        \hline
    \end{tabular}
    \caption{The redesigned version of the question in Figure~\ref{fig:cs1-question} incorporates GenAI to emphasise concepts and processes rather than product.}
    \label{fig:cs1-redesigned}
\end{figure}

The focus of the redesigned assignments is to assist with learning rather than to evaluate learning. 
Although we believe that GenAI should be used in assignments, we also believe it is beneficial for courses to include components of secure assessments (in-person, invigilated, and GenAI-prohibited) to ensure that we measure student capability for accreditation purposes and to focus attention on that capability rather than relying on an outside source (such as GenAI) to produce outputs that students are expected to complete themselves.  We enforce secure assessments by having students submit their code on a web-based automated marking tool, such as Coderunner~\cite{lobb2016coderunner}, and a firewall that only allows Internet access to the said marking tool. This eliminates the possibility of students using GenAI during secure assessment components.

A potential challenge arising from more qualitative assignments, or focusing on process rather than product, is the increased workload for educators to grade student submissions. Using the traditional grading method of manually reading submitted material may be ineffective, and innovative methods may need to be devised to speed up the grading process. One possibility is to design a proxy for students to access GenAI tools that store interactions that may be used to automate grading purposes. 
The requirements of the grading process may vary between assignments, but they should be considered when designing the assignments.

The redesign of assignments also raises concerns regarding the correctness of responses generated by GenAI since students risk counterproductive learning if the generated information is incorrect or misleading. However, instead of perceiving this as counterproductive learning, educators must teach students to identify incorrect information by fact-checking using trusted non-AI-generated resources such as textbooks and teaching staff, which is an essential skill in a discipline where much of the recent content knowledge is acquired through community sources (such as forums, blog posts, and online documentation). This can strengthen student understanding of topics and simultaneously improve their metacognitive skills.

\subsection{Strategy 2: Teach Metacognition}
\label{metacognition}
With increasing online resources, students are becoming less reliant on educators, as their first source of study assistance is usually the Internet~\cite{hou2024effects} because Internet searches supply answers instantly. One main advantage of educators' responses to student queries is that they provide personalised feedback that answer specific details of the questions that online searches may not supply. However, now GenAI tools are easily accessible and students can receive unlimited, immediate, and personalised feedback.  Consequentially, we have observed fewer help requests from students in our computing laboratories, as described in Subsection~\ref{queries}.  This indicates that we may better support student learning by increasing attention to skills students cannot learn simply with assistance from GenAI, such as metacognitive skills crucial when learning CS~\cite{prather2020we}.

In Subsection~\ref{queries}, we outline the differences in the approaches of using GenAI between more motivated students and those less motivated, which may also be attributed to the differences in metacognitive skills, where more motivated students understand the harmful effects of using GenAI without engaging in learning. In contrast, those less motivated may not consider the future effects of copying from GenAI, such as the lack of understanding leading to difficulties during assessments. We believe that it should be the educators' responsibility to highlight the risks of copying from GenAI and the correct and most effective ways of using GenAI when learning, such as asking GenAI to refrain from providing full solutions and instead provide explanations and examples.

\citet{kazemitabaar2024exploringdesignspacecognitive} present a variety of GenAI-supported tasks that align with different levels of Bloom's taxonomy.  These tasks require tools designed to scaffold learning with GenAI and could be introduced into CS classrooms with guidance that raises awareness of how the activities impact learning and the role of GenAI in that process.  

\begin{figure}
    \centering
    \begin{tabular}{|p{0.45\textwidth}|}
        \hline
        \textit{Prompt} \\
        How to write a function to check whether a number is prime? \\
        \hline
    \end{tabular}
    \begin{tabular}{|p{0.45\textwidth}|}
        \hline
        \textit{Response} \\
        1. Special Cases ... \\
        2. Divisibility Check ... \\
        \textit{[FULL CODE SOLUTION WITH LIMITED COMMENTS]} \\
        \hline
    \end{tabular}
    \caption{A general prompt asking a CS1 question and an excerpt of the corresponding response from ChatGPT}
    \label{fig:prompt-general}
\end{figure}

\begin{figure}
    \centering
    \begin{tabular}{|p{0.45\textwidth}|}
        \hline
        \textit{Prompt} \\
        Write a function to check whether a number is prime. Please do not include any code and only explain the algorithm in detail. \\
        \hline
    \end{tabular}
    \begin{tabular}{|p{0.45\textwidth}|}
        \hline
        \textit{Response} \\
        1. Handle Small Numbers ... \\
        2. Eliminate Even and Small Odd Divisors Early ... \\
        3. Check for Divisors up to the Square Root of $n$ ... \\
        4. Terminate if Any Divisor is Found ... \\
        \hline
    \end{tabular}
    \caption{An effective prompt (requesting for no code and explanations only) asking a CS1 question and an excerpt of the corresponding response from ChatGPT}
    \label{fig:prompt-explanations}
\end{figure}

\begin{figure}
    \centering
    \begin{tabular}{|p{0.45\textwidth}|}
        \hline
        \textit{Prompt} \\
        Write a function to check whether a number is prime. Please provide more comments in the code and explain how each line works. \\
        \hline
    \end{tabular}
    \begin{tabular}{|p{0.45\textwidth}|}
        \hline
        \textit{Response} \\
        1. Function Definition ... \\
        2. Check if n is less than or equal to 1 ... \\
        3. Loop from 2 to the square root of n ... \\
        4. Check if n is divisible by any number in the range ... \\
        5. Return True if no factors are found ... \\
        \textit{[FULL CODE SOLUTION WITH DETAILED COMMENTS]} \\
        \hline
    \end{tabular}
    \caption{An effective prompt (requesting for further details) asking a CS1 question and an excerpt of the corresponding response from ChatGPT}
    \label{fig:prompt-comments}
\end{figure}

More generally available tools, such as ChatGPT, are more likely to be used in practice by students working independently on programming tasks.  The way students assemble their prompts significantly influences the content, thus usefulness, of the responses, so educators should teach students about the characteristics of effective prompts for students. Teachers can easily demonstrate how to prompt ChatGPT and the impact of different prompts.  The Prompt Problems exercise described in Subsection~\ref{promptproblems} illustrates how guidance about prompts could subsequently be assessed as part of a course activity.

For example, in the scenario of a CS1 student attempting to understand the process of writing a prime checking function with the assistance of ChatGPT, Figure~\ref{fig:prompt-general} shows an example of a general prompt, where the explanation lacks details and may be difficult to understand for a novice programmer. Additionally, it provides the full model solution, which reduces the learning that occurs as students explore the solution space. In contrast, Figure~\ref{fig:prompt-explanations} shows an example of a prompt requesting no code and only explanations. This is more valuable for learning than the general prompt, as the students can then attempt to write the code themselves from the information generated by ChatGPT. If students fail to write the code themselves, they can use the prompt shown in Figure~\ref{fig:prompt-comments}, which expects a full code solution but with further comments and details. As expected, the corresponding explanation and code are more comprehensive and explicit than that shown in Figure~\ref{fig:prompt-general}, even providing explanations for the use of the if-statements and for-loops.
Using the prompts shown in Figure~\ref{fig:prompt-explanations} and Figure~\ref{fig:prompt-comments}, 
and observing the corresponding annotated outputs, may help a student develop a deeper understanding of the problem and the computational steps involved, compared with only using the prompt shown in Figure~\ref{fig:prompt-general}.


The characteristics of effective prompts can vary across courses and disciplines. Therefore, educators should provide course-specific guidance on crafting good prompts for queries within the context of their course. Additionally, students need to recognize that GenAI responses are not always accurate, and they must develop the ability to identify misinformation. This process can serve as both a valuable learning experience and an important metacognitive skill.

AI-generated content comes in various forms, each potentially requiring different fact-checking methods. Educators should demonstrate the fact-checking techniques most relevant to their courses to teach students how to identify misinformation effectively. For example, educators can present snippets of AI-generated content and walk students through the process of verifying this information using course materials, textbooks, documentation, and trusted websites. Similarly, educators can demonstrate how to test AI-generated code for correctness by running it. If the code produces incorrect results, educators can showcase the debugging process, which is an essential skill, particularly as code generation becomes more prevalent.

\subsection{Future Trends}

The CE landscape will inevitably undergo further significant change due to GenAI. While many new teaching practices have emerged in response to GenAI, such as those outlined in this work, we argue that further research is essential to evaluate their validity and effectiveness. This includes trialling innovative teaching methods, gathering student feedback, and assessing student performance. It is critical to thoroughly investigate the concrete impacts of GenAI and proactively prepare for these changes before the research gap becomes too wide. Neglecting this need could result in reduced student productivity due to ineffective GenAI usage and leave graduates ill-prepared because of outdated teaching practices.

We believe that future graduates must be proficient and experienced in leveraging GenAI to support their work. Therefore, methods for effectively using GenAI should be integrated into teaching curricula. We urge the CE community to prioritize research that examines the tangible impacts of GenAI, including its influence on course structure, content, and teaching methods. Such research will enable educators to make well-informed decisions as we adapt our teaching practices to the evolving role of GenAI in education.

\section{Conclusion}
In this paper, we shared our experiences with integrating GenAI into CS-focused educational settings and the adjustments we have made to our teaching practices in recent years following the popularization of GenAI. Based on these experiences, we emphasized the need to redefine the role of CS educators in the GenAI era, focusing more on developing students' metacognitive skills.
To support this shift, we proposed two strategies: redesigning take-home assignments to incorporate GenAI use and adapting teaching approaches to emphasize metacognition by demonstrating the correct and effective use of GenAI. Additionally, we highlighted the importance of conducting further research to evaluate the concrete impacts of GenAI on computing education, particularly in assessing the validity and effectiveness of emerging teaching practices.
We hope this paper provides valuable insights and inspiration for navigating the evolving landscape of CS education as GenAI continues to advance.


\balance
\bibliographystyle{ACM-Reference-Format}
\bibliography{ref}


\begin{thebibliography}{44}


\ifx \showCODEN    \undefined \def \showCODEN     #1{\unskip}     \fi
\ifx \showDOI      \undefined \def \showDOI       #1{#1}\fi
\ifx \showISBNx    \undefined \def \showISBNx     #1{\unskip}     \fi
\ifx \showISBNxiii \undefined \def \showISBNxiii  #1{\unskip}     \fi
\ifx \showISSN     \undefined \def \showISSN      #1{\unskip}     \fi
\ifx \showLCCN     \undefined \def \showLCCN      #1{\unskip}     \fi
\ifx \shownote     \undefined \def \shownote      #1{#1}          \fi
\ifx \showarticletitle \undefined \def \showarticletitle #1{#1}   \fi
\ifx \showURL      \undefined \def \showURL       {\relax}        \fi
\providecommand\bibfield[2]{#2}
\providecommand\bibinfo[2]{#2}
\providecommand\natexlab[1]{#1}
\providecommand\showeprint[2][]{arXiv:#2}

\bibitem[Becker et~al\mbox{.}(2023)]%
        {becker2023programming}
\bibfield{author}{\bibinfo{person}{Brett~A Becker}, \bibinfo{person}{Paul Denny}, \bibinfo{person}{James Finnie-Ansley}, \bibinfo{person}{Andrew Luxton-Reilly}, \bibinfo{person}{James Prather}, {and} \bibinfo{person}{Eddie~Antonio Santos}.} \bibinfo{year}{2023}\natexlab{}.
\newblock \showarticletitle{Programming is hard-or at least it used to be: Educational opportunities and challenges of ai code generation}. In \bibinfo{booktitle}{\emph{Proceedings of the 54th ACM Technical Symposium on Computer Science Education V. 1}}. \bibinfo{pages}{500--506}.
\newblock


\bibitem[Bernstein et~al\mbox{.}(2024)]%
        {bernstein2024nesting}
\bibfield{author}{\bibinfo{person}{Seth Bernstein}, \bibinfo{person}{Paul Denny}, \bibinfo{person}{Juho Leinonen}, \bibinfo{person}{Lauren Kan}, \bibinfo{person}{Arto Hellas}, \bibinfo{person}{Matt Littlefield}, \bibinfo{person}{Sami Sarsa}, {and} \bibinfo{person}{Stephen Macneil}.} \bibinfo{year}{2024}\natexlab{}.
\newblock \showarticletitle{"Like a Nesting Doll": Analyzing Recursion Analogies Generated by CS Students Using Large Language Models}. In \bibinfo{booktitle}{\emph{Proceedings of the 2024 on Innovation and Technology in Computer Science Education V. 1}} (Milan, Italy) \emph{(\bibinfo{series}{ITiCSE 2024})}. \bibinfo{publisher}{Association for Computing Machinery}, \bibinfo{address}{New York, NY, USA}, \bibinfo{pages}{122–128}.
\newblock
\showISBNx{9798400706004}
\urldef\tempurl%
\url{https://doi.org/10.1145/3649217.3653533}
\showDOI{\tempurl}


\bibitem[Cao and Dede(2023)]%
        {cao2023navigating}
\bibfield{author}{\bibinfo{person}{Lydia Cao} {and} \bibinfo{person}{Chris Dede}.} \bibinfo{year}{2023}\natexlab{}.
\newblock \showarticletitle{Navigating a world of generative AI: Suggestions for educators}.
\newblock \bibinfo{journal}{\emph{The Next Level Lab at Harvard Graduate School of Education. President and Fellows of Harvard College: Cambridge}} (\bibinfo{year}{2023}).
\newblock


\bibitem[del Carpio~Gutierrez et~al\mbox{.}(2024)]%
        {gutierrez2024automating}
\bibfield{author}{\bibinfo{person}{Andre del Carpio~Gutierrez}, \bibinfo{person}{Paul Denny}, {and} \bibinfo{person}{Andrew Luxton-Reilly}.} \bibinfo{year}{2024}\natexlab{}.
\newblock \showarticletitle{Automating Personalized Parsons Problems with Customized Contexts and Concepts}. In \bibinfo{booktitle}{\emph{Proceedings of the 2024 on Innovation and Technology in Computer Science Education V. 1}} (Milan, Italy) \emph{(\bibinfo{series}{ITiCSE 2024})}. \bibinfo{publisher}{Association for Computing Machinery}, \bibinfo{address}{New York, NY, USA}, \bibinfo{pages}{688–694}.
\newblock
\showISBNx{9798400706004}
\urldef\tempurl%
\url{https://doi.org/10.1145/3649217.3653568}
\showDOI{\tempurl}


\bibitem[Denny et~al\mbox{.}(2023)]%
        {denny2023conversing}
\bibfield{author}{\bibinfo{person}{Paul Denny}, \bibinfo{person}{Viraj Kumar}, {and} \bibinfo{person}{Nasser Giacaman}.} \bibinfo{year}{2023}\natexlab{}.
\newblock \showarticletitle{Conversing with copilot: Exploring prompt engineering for solving {CS1} problems using natural language}. In \bibinfo{booktitle}{\emph{Proceedings of the 54th ACM Technical Symposium on Computer Science Education V. 1}}. \bibinfo{pages}{1136--1142}.
\newblock


\bibitem[Denny et~al\mbox{.}(2024a)]%
        {denny2024prompt}
\bibfield{author}{\bibinfo{person}{Paul Denny}, \bibinfo{person}{Juho Leinonen}, \bibinfo{person}{James Prather}, \bibinfo{person}{Andrew Luxton-Reilly}, \bibinfo{person}{Thezyrie Amarouche}, \bibinfo{person}{Brett~A. Becker}, {and} \bibinfo{person}{Brent~N. Reeves}.} \bibinfo{year}{2024}\natexlab{a}.
\newblock \showarticletitle{Prompt Problems: A New Programming Exercise for the Generative AI Era}. In \bibinfo{booktitle}{\emph{Proceedings of the 55th ACM Technical Symposium on Computer Science Education V. 1}} (Portland, OR, USA) \emph{(\bibinfo{series}{SIGCSE 2024})}. \bibinfo{publisher}{Association for Computing Machinery}, \bibinfo{address}{New York, NY, USA}, \bibinfo{pages}{296–302}.
\newblock
\showISBNx{9798400704239}
\urldef\tempurl%
\url{https://doi.org/10.1145/3626252.3630909}
\showDOI{\tempurl}


\bibitem[Denny et~al\mbox{.}(2024b)]%
        {denny2024desirable}
\bibfield{author}{\bibinfo{person}{Paul Denny}, \bibinfo{person}{Stephen MacNeil}, \bibinfo{person}{Jaromir Savelka}, \bibinfo{person}{Leo Porter}, {and} \bibinfo{person}{Andrew Luxton-Reilly}.} \bibinfo{year}{2024}\natexlab{b}.
\newblock \showarticletitle{Desirable Characteristics for AI Teaching Assistants in Programming Education}. In \bibinfo{booktitle}{\emph{Proceedings of the 2024 on Innovation and Technology in Computer Science Education V. 1}} (Milan, Italy) \emph{(\bibinfo{series}{ITiCSE 2024})}. \bibinfo{publisher}{Association for Computing Machinery}, \bibinfo{address}{New York, NY, USA}, \bibinfo{pages}{408–414}.
\newblock
\showISBNx{9798400706004}
\urldef\tempurl%
\url{https://doi.org/10.1145/3649217.3653574}
\showDOI{\tempurl}


\bibitem[Denny et~al\mbox{.}(2024c)]%
        {denny2024cacm}
\bibfield{author}{\bibinfo{person}{Paul Denny}, \bibinfo{person}{James Prather}, \bibinfo{person}{Brett~A. Becker}, \bibinfo{person}{James Finnie-Ansley}, \bibinfo{person}{Arto Hellas}, \bibinfo{person}{Juho Leinonen}, \bibinfo{person}{Andrew Luxton-Reilly}, \bibinfo{person}{Brent~N. Reeves}, \bibinfo{person}{Eddie~Antonio Santos}, {and} \bibinfo{person}{Sami Sarsa}.} \bibinfo{year}{2024}\natexlab{c}.
\newblock \showarticletitle{Computing Education in the Era of Generative AI}.
\newblock \bibinfo{journal}{\emph{Commun. ACM}} \bibinfo{volume}{67}, \bibinfo{number}{2} (\bibinfo{date}{Jan.} \bibinfo{year}{2024}), \bibinfo{pages}{56–67}.
\newblock
\showISSN{0001-0782}
\urldef\tempurl%
\url{https://doi.org/10.1145/3624720}
\showDOI{\tempurl}


\bibitem[Feng et~al\mbox{.}(2024a)]%
        {feng2024more}
\bibfield{author}{\bibinfo{person}{Tony~Haoran Feng}, \bibinfo{person}{Paul Denny}, \bibinfo{person}{Burkhard Wuensche}, \bibinfo{person}{Andrew Luxton-Reilly}, {and} \bibinfo{person}{Steffan Hooper}.} \bibinfo{year}{2024}\natexlab{a}.
\newblock \showarticletitle{More Than Meets the AI: Evaluating the performance of GPT-4 on Computer Graphics assessment questions}. In \bibinfo{booktitle}{\emph{Proceedings of the 26th Australasian Computing Education Conference}}. \bibinfo{pages}{182--191}.
\newblock


\bibitem[Feng et~al\mbox{.}(2024b)]%
        {feng2024eye}
\bibfield{author}{\bibinfo{person}{Tony~Haoran Feng}, \bibinfo{person}{Paul Denny}, \bibinfo{person}{Burkhard~C W{\"u}nsche}, \bibinfo{person}{Andrew Luxton-Reilly}, {and} \bibinfo{person}{Jacqueline Whalley}.} \bibinfo{year}{2024}\natexlab{b}.
\newblock \showarticletitle{An Eye for an AI: Evaluating GPT-4o's Visual Perception Skills and Geometric Reasoning Skills Using Computer Graphics Questions}. In \bibinfo{booktitle}{\emph{SIGGRAPH Asia 2024 Educator's Forum}}. \bibinfo{pages}{1--8}.
\newblock


\bibitem[Feng et~al\mbox{.}(2024c)]%
        {feng2024can}
\bibfield{author}{\bibinfo{person}{Tony~Haoran Feng}, \bibinfo{person}{Burkhard~C. Wünsche}, \bibinfo{person}{Paul Denny}, \bibinfo{person}{Andrew Luxton-Reilly}, {and} \bibinfo{person}{Steffan Hooper}.} \bibinfo{year}{2024}\natexlab{c}.
\newblock \showarticletitle{{Can GPT-4 Trace Rays}}. In \bibinfo{booktitle}{\emph{Eurographics 2024 - Education Papers}}, \bibfield{editor}{\bibinfo{person}{Beatriz Sousa~Santos} {and} \bibinfo{person}{Eike Anderson}} (Eds.). \bibinfo{publisher}{The Eurographics Association}.
\newblock
\showISBNx{978-3-03868-238-7}
\showISSN{1017-4656}
\urldef\tempurl%
\url{https://doi.org/10.2312/eged.20241003}
\showDOI{\tempurl}


\bibitem[Finnie-Ansley et~al\mbox{.}(2022)]%
        {finnie2022robots}
\bibfield{author}{\bibinfo{person}{James Finnie-Ansley}, \bibinfo{person}{Paul Denny}, \bibinfo{person}{Brett~A Becker}, \bibinfo{person}{Andrew Luxton-Reilly}, {and} \bibinfo{person}{James Prather}.} \bibinfo{year}{2022}\natexlab{}.
\newblock \showarticletitle{The robots are coming: Exploring the implications of openai codex on introductory programming}. In \bibinfo{booktitle}{\emph{Proceedings of the 24th Australasian Computing Education Conference}}. \bibinfo{publisher}{Association for Computing Machinery}, \bibinfo{pages}{10--19}.
\newblock


\bibitem[Hooper et~al\mbox{.}(2024)]%
        {hooper2024advancing}
\bibfield{author}{\bibinfo{person}{Steffan Hooper}, \bibinfo{person}{Burkhard~C W{\"u}nsche}, \bibinfo{person}{Andrew Luxton-Reilly}, \bibinfo{person}{Paul Denny}, {and} \bibinfo{person}{Tony~Haoran Feng}.} \bibinfo{year}{2024}\natexlab{}.
\newblock \showarticletitle{Advancing Automated Assessment Tools-Opportunities for Innovations in Upper-level Computing Courses: A Position Paper}. In \bibinfo{booktitle}{\emph{Proceedings of the 55th ACM Technical Symposium on Computer Science Education V. 1}}. \bibinfo{pages}{519--525}.
\newblock


\bibitem[Hou et~al\mbox{.}(2024a)]%
        {hou2024effects}
\bibfield{author}{\bibinfo{person}{Irene Hou}, \bibinfo{person}{Sophia Mettille}, \bibinfo{person}{Owen Man}, \bibinfo{person}{Zhuo Li}, \bibinfo{person}{Cynthia Zastudil}, {and} \bibinfo{person}{Stephen MacNeil}.} \bibinfo{year}{2024}\natexlab{a}.
\newblock \showarticletitle{The Effects of Generative AI on Computing Students’ Help-Seeking Preferences}. In \bibinfo{booktitle}{\emph{Proceedings of the 26th Australasian Computing Education Conference}}. \bibinfo{pages}{39--48}.
\newblock


\bibitem[Hou et~al\mbox{.}(2024b)]%
        {hou2024tailor}
\bibfield{author}{\bibinfo{person}{Xinying Hou}, \bibinfo{person}{Zihan Wu}, \bibinfo{person}{Xu Wang}, {and} \bibinfo{person}{Barbara~J. Ericson}.} \bibinfo{year}{2024}\natexlab{b}.
\newblock \showarticletitle{CodeTailor: LLM-Powered Personalized Parsons Puzzles for Engaging Support While Learning Programming}. In \bibinfo{booktitle}{\emph{Proceedings of the Eleventh ACM Conference on Learning @ Scale}} (Atlanta, GA, USA) \emph{(\bibinfo{series}{L@S '24})}. \bibinfo{publisher}{Association for Computing Machinery}, \bibinfo{address}{New York, NY, USA}, \bibinfo{pages}{51–62}.
\newblock
\showISBNx{9798400706332}
\urldef\tempurl%
\url{https://doi.org/10.1145/3657604.3662032}
\showDOI{\tempurl}


\bibitem[Indriasari et~al\mbox{.}(2020)]%
        {indriasari2020review}
\bibfield{author}{\bibinfo{person}{Theresia~Devi Indriasari}, \bibinfo{person}{Andrew Luxton-Reilly}, {and} \bibinfo{person}{Paul Denny}.} \bibinfo{year}{2020}\natexlab{}.
\newblock \showarticletitle{A Review of Peer Code Review in Higher Education}.
\newblock \bibinfo{journal}{\emph{ACM Trans. Comput. Educ.}} \bibinfo{volume}{20}, \bibinfo{number}{3}, Article \bibinfo{articleno}{22} (\bibinfo{date}{Sept.} \bibinfo{year}{2020}), \bibinfo{numpages}{25}~pages.
\newblock
\urldef\tempurl%
\url{https://doi.org/10.1145/3403935}
\showDOI{\tempurl}


\bibitem[Jury et~al\mbox{.}(2024)]%
        {jury2024evaluating}
\bibfield{author}{\bibinfo{person}{Breanna Jury}, \bibinfo{person}{Angela Lorusso}, \bibinfo{person}{Juho Leinonen}, \bibinfo{person}{Paul Denny}, {and} \bibinfo{person}{Andrew Luxton-Reilly}.} \bibinfo{year}{2024}\natexlab{}.
\newblock \showarticletitle{Evaluating LLM-generated Worked Examples in an Introductory Programming Course}. In \bibinfo{booktitle}{\emph{Proceedings of the 26th Australasian Computing Education Conference}} (Sydney, NSW, Australia) \emph{(\bibinfo{series}{ACE '24})}. \bibinfo{publisher}{Association for Computing Machinery}, \bibinfo{address}{New York, NY, USA}, \bibinfo{pages}{77–86}.
\newblock
\showISBNx{9798400716195}
\urldef\tempurl%
\url{https://doi.org/10.1145/3636243.3636252}
\showDOI{\tempurl}


\bibitem[Kazemitabaar et~al\mbox{.}(2024a)]%
        {kazemitabaar2024exploringdesignspacecognitive}
\bibfield{author}{\bibinfo{person}{Majeed Kazemitabaar}, \bibinfo{person}{Oliver Huang}, \bibinfo{person}{Sangho Suh}, \bibinfo{person}{Austin~Z. Henley}, {and} \bibinfo{person}{Tovi Grossman}.} \bibinfo{year}{2024}\natexlab{a}.
\newblock \bibinfo{title}{Exploring the Design Space of Cognitive Engagement Techniques with AI-Generated Code for Enhanced Learning}.
\newblock
\newblock
\showeprint[arxiv]{2410.08922}~[cs.HC]
\urldef\tempurl%
\url{https://arxiv.org/abs/2410.08922}
\showURL{%
\tempurl}


\bibitem[Kazemitabaar et~al\mbox{.}(2024b)]%
        {kazemitabaar2024codeaid}
\bibfield{author}{\bibinfo{person}{Majeed Kazemitabaar}, \bibinfo{person}{Runlong Ye}, \bibinfo{person}{Xiaoning Wang}, \bibinfo{person}{Austin~Z Henley}, \bibinfo{person}{Paul Denny}, \bibinfo{person}{Michelle Craig}, {and} \bibinfo{person}{Tovi Grossman}.} \bibinfo{year}{2024}\natexlab{b}.
\newblock \showarticletitle{CodeAid: Evaluating a Classroom Deployment of an LLM-based Programming Assistant that Balances Student and Educator Needs}.
\newblock \bibinfo{journal}{\emph{arXiv preprint arXiv:2401.11314}} (\bibinfo{year}{2024}).
\newblock


\bibitem[Kerslake et~al\mbox{.}(2024)]%
        {kerslake2024integrating}
\bibfield{author}{\bibinfo{person}{Chris Kerslake}, \bibinfo{person}{Paul Denny}, \bibinfo{person}{David~H. Smith}, \bibinfo{person}{James Prather}, \bibinfo{person}{Juho Leinonen}, \bibinfo{person}{Andrew Luxton-Reilly}, {and} \bibinfo{person}{Stephen MacNeil}.} \bibinfo{year}{2024}\natexlab{}.
\newblock \showarticletitle{Integrating Natural Language Prompting Tasks in Introductory Programming Courses}. In \bibinfo{booktitle}{\emph{Proceedings of the 2024 on ACM Virtual Global Computing Education Conference V. 1}} (Virtual Event, NC, USA) \emph{(\bibinfo{series}{SIGCSE Virtual 2024})}. \bibinfo{publisher}{Association for Computing Machinery}, \bibinfo{address}{New York, NY, USA}, \bibinfo{pages}{88–94}.
\newblock
\showISBNx{9798400705984}
\urldef\tempurl%
\url{https://doi.org/10.1145/3649165.3690125}
\showDOI{\tempurl}


\bibitem[{Khan Academy}(2024)]%
        {gpt4o_khan}
\bibfield{author}{\bibinfo{person}{{Khan Academy}}.} \bibinfo{year}{2024}\natexlab{}.
\newblock \bibinfo{title}{{GPT-4o (Omni) math tutoring demo on Khan Academy - YouTube}}.
\newblock \bibinfo{howpublished}{\url{https://www.youtube.com/watch?v=IvXZCocyU_M}}.
\newblock
\newblock
\shownote{[Accessed 22-05-2024]}.


\bibitem[Kumar et~al\mbox{.}(2024)]%
        {kumar2024computer}
\bibfield{author}{\bibinfo{person}{Amruth~N. Kumar}, \bibinfo{person}{Rajendra~K. Raj}, \bibinfo{person}{Sherif~G. Aly}, \bibinfo{person}{Monica~D. Anderson}, \bibinfo{person}{Brett~A. Becker}, \bibinfo{person}{Richard~L. Blumenthal}, \bibinfo{person}{Eric Eaton}, \bibinfo{person}{Susan~L. Epstein}, \bibinfo{person}{Michael Goldweber}, \bibinfo{person}{Pankaj Jalote}, \bibinfo{person}{Douglas Lea}, \bibinfo{person}{Michael Oudshoorn}, \bibinfo{person}{Marcelo Pias}, \bibinfo{person}{Susan Reiser}, \bibinfo{person}{Christian Servin}, \bibinfo{person}{Rahul Simha}, \bibinfo{person}{Titus Winters}, {and} \bibinfo{person}{Qiao Xiang}.} \bibinfo{year}{2024}\natexlab{}.
\newblock \bibinfo{booktitle}{\emph{Computer Science Curricula 2023}}.
\newblock \bibinfo{publisher}{Association for Computing Machinery}, \bibinfo{address}{New York, NY, USA}.
\newblock
\showISBNx{9798400710339}


\bibitem[Lekshmi-Narayanan et~al\mbox{.}(2024)]%
        {lekshminarayanan2024explainingcodeexamplesintroductory}
\bibfield{author}{\bibinfo{person}{Arun-Balajiee Lekshmi-Narayanan}, \bibinfo{person}{Priti Oli}, \bibinfo{person}{Jeevan Chapagain}, \bibinfo{person}{Mohammad Hassany}, \bibinfo{person}{Rabin Banjade}, \bibinfo{person}{Peter Brusilovsky}, {and} \bibinfo{person}{Vasile Rus}.} \bibinfo{year}{2024}\natexlab{}.
\newblock \bibinfo{title}{Explaining Code Examples in Introductory Programming Courses: LLM vs Humans}.
\newblock
\newblock
\showeprint[arxiv]{2403.05538}~[cs.CY]
\urldef\tempurl%
\url{https://arxiv.org/abs/2403.05538}
\showURL{%
\tempurl}


\bibitem[Lobb and Harlow(2016)]%
        {lobb2016coderunner}
\bibfield{author}{\bibinfo{person}{Richard Lobb} {and} \bibinfo{person}{Jenny Harlow}.} \bibinfo{year}{2016}\natexlab{}.
\newblock \showarticletitle{Coderunner: A tool for assessing computer programming skills}.
\newblock \bibinfo{journal}{\emph{ACM Inroads}} \bibinfo{volume}{7}, \bibinfo{number}{1} (\bibinfo{year}{2016}), \bibinfo{pages}{47--51}.
\newblock


\bibitem[Lye and Lim(2024)]%
        {lye2024generative}
\bibfield{author}{\bibinfo{person}{Che~Yee Lye} {and} \bibinfo{person}{Lyndon Lim}.} \bibinfo{year}{2024}\natexlab{}.
\newblock \showarticletitle{Generative Artificial Intelligence in Tertiary Education: Assessment Redesign Principles and Considerations}.
\newblock \bibinfo{journal}{\emph{Education Sciences}} \bibinfo{volume}{14}, \bibinfo{number}{6} (\bibinfo{year}{2024}), \bibinfo{pages}{569}.
\newblock


\bibitem[Ma et~al\mbox{.}(2024)]%
        {ma2024hypocompass}
\bibfield{author}{\bibinfo{person}{Qianou Ma}, \bibinfo{person}{Hua Shen}, \bibinfo{person}{Kenneth Koedinger}, {and} \bibinfo{person}{Sherry~Tongshuang Wu}.} \bibinfo{year}{2024}\natexlab{}.
\newblock \showarticletitle{How to Teach Programming in the AI Era? Using LLMs as a Teachable Agent for Debugging}. In \bibinfo{booktitle}{\emph{Artificial Intelligence in Education}}, \bibfield{editor}{\bibinfo{person}{Andrew~M. Olney}, \bibinfo{person}{Irene-Angelica Chounta}, \bibinfo{person}{Zitao Liu}, \bibinfo{person}{Olga~C. Santos}, {and} \bibinfo{person}{Ig~Ibert Bittencourt}} (Eds.). \bibinfo{publisher}{Springer Nature Switzerland}, \bibinfo{address}{Cham}, \bibinfo{pages}{265--279}.
\newblock
\showISBNx{978-3-031-64302-6}


\bibitem[MacNeil et~al\mbox{.}(2023)]%
        {macneil2023experiences}
\bibfield{author}{\bibinfo{person}{Stephen MacNeil}, \bibinfo{person}{Andrew Tran}, \bibinfo{person}{Arto Hellas}, \bibinfo{person}{Joanne Kim}, \bibinfo{person}{Sami Sarsa}, \bibinfo{person}{Paul Denny}, \bibinfo{person}{Seth Bernstein}, {and} \bibinfo{person}{Juho Leinonen}.} \bibinfo{year}{2023}\natexlab{}.
\newblock \showarticletitle{Experiences from Using Code Explanations Generated by Large Language Models in a Web Software Development E-Book}. In \bibinfo{booktitle}{\emph{Proceedings of the 54th ACM Technical Symposium on Computer Science Education V. 1}} (Toronto ON, Canada) \emph{(\bibinfo{series}{SIGCSE 2023})}. \bibinfo{publisher}{Association for Computing Machinery}, \bibinfo{address}{New York, NY, USA}, \bibinfo{pages}{931–937}.
\newblock
\showISBNx{9781450394314}
\urldef\tempurl%
\url{https://doi.org/10.1145/3545945.3569785}
\showDOI{\tempurl}


\bibitem[Milano et~al\mbox{.}(2023)]%
        {milano2023large}
\bibfield{author}{\bibinfo{person}{Silvia Milano}, \bibinfo{person}{Joshua~A McGrane}, {and} \bibinfo{person}{Sabina Leonelli}.} \bibinfo{year}{2023}\natexlab{}.
\newblock \showarticletitle{Large language models challenge the future of higher education}.
\newblock \bibinfo{journal}{\emph{Nature Machine Intelligence}} \bibinfo{volume}{5}, \bibinfo{number}{4} (\bibinfo{year}{2023}), \bibinfo{pages}{333--334}.
\newblock


\bibitem[{OpenAI}(2024)]%
        {chatgpt}
\bibfield{author}{\bibinfo{person}{{OpenAI}}.} \bibinfo{year}{2024}\natexlab{}.
\newblock \bibinfo{title}{{ChatGPT}}.
\newblock \bibinfo{howpublished}{\url{https://chatgpt.com/}}.
\newblock
\newblock
\shownote{[Accessed 10-12-2024]}.


\bibitem[Orenstrakh et~al\mbox{.}(2023)]%
        {orenstrakh2023detecting}
\bibfield{author}{\bibinfo{person}{Michael~Sheinman Orenstrakh}, \bibinfo{person}{Oscar Karnalim}, \bibinfo{person}{Carlos~Anibal Suarez}, {and} \bibinfo{person}{Michael Liut}.} \bibinfo{year}{2023}\natexlab{}.
\newblock \showarticletitle{Detecting llm-generated text in computing education: A comparative study for chatgpt cases}.
\newblock \bibinfo{journal}{\emph{arXiv preprint arXiv:2307.07411}} (\bibinfo{year}{2023}).
\newblock


\bibitem[Perkins(2023)]%
        {perkins2023academic}
\bibfield{author}{\bibinfo{person}{Mike Perkins}.} \bibinfo{year}{2023}\natexlab{}.
\newblock \showarticletitle{Academic Integrity considerations of AI Large Language Models in the post-pandemic era: ChatGPT and beyond}.
\newblock \bibinfo{journal}{\emph{Journal of university teaching \& learning practice}} \bibinfo{volume}{20}, \bibinfo{number}{2} (\bibinfo{year}{2023}), \bibinfo{pages}{07}.
\newblock


\bibitem[Prather et~al\mbox{.}(2020)]%
        {prather2020we}
\bibfield{author}{\bibinfo{person}{James Prather}, \bibinfo{person}{Brett~A Becker}, \bibinfo{person}{Michelle Craig}, \bibinfo{person}{Paul Denny}, \bibinfo{person}{Dastyni Loksa}, {and} \bibinfo{person}{Lauren Margulieux}.} \bibinfo{year}{2020}\natexlab{}.
\newblock \showarticletitle{What do we think we think we are doing? Metacognition and self-regulation in programming}. In \bibinfo{booktitle}{\emph{Proceedings of the 2020 ACM conference on international computing education research}}. \bibinfo{pages}{2--13}.
\newblock


\bibitem[Prather et~al\mbox{.}(2023)]%
        {prather2023robots}
\bibfield{author}{\bibinfo{person}{James Prather}, \bibinfo{person}{Paul Denny}, \bibinfo{person}{Juho Leinonen}, \bibinfo{person}{Brett~A Becker}, \bibinfo{person}{Ibrahim Albluwi}, \bibinfo{person}{Michelle Craig}, \bibinfo{person}{Hieke Keuning}, \bibinfo{person}{Natalie Kiesler}, \bibinfo{person}{Tobias Kohn}, \bibinfo{person}{Andrew Luxton-Reilly}, {et~al\mbox{.}}} \bibinfo{year}{2023}\natexlab{}.
\newblock \showarticletitle{The robots are here: Navigating the generative ai revolution in computing education}.
\newblock In \bibinfo{booktitle}{\emph{Proceedings of the 2023 Working Group Reports on Innovation and Technology in Computer Science Education}}. \bibinfo{publisher}{Association for Computing Machinery}, \bibinfo{pages}{108--159}.
\newblock


\bibitem[Prather et~al\mbox{.}(2024)]%
        {prather2024widening}
\bibfield{author}{\bibinfo{person}{James Prather}, \bibinfo{person}{Brent~N Reeves}, \bibinfo{person}{Juho Leinonen}, \bibinfo{person}{Stephen MacNeil}, \bibinfo{person}{Arisoa~S Randrianasolo}, \bibinfo{person}{Brett~A. Becker}, \bibinfo{person}{Bailey Kimmel}, \bibinfo{person}{Jared Wright}, {and} \bibinfo{person}{Ben Briggs}.} \bibinfo{year}{2024}\natexlab{}.
\newblock \showarticletitle{The Widening Gap: The Benefits and Harms of Generative AI for Novice Programmers}. In \bibinfo{booktitle}{\emph{Proceedings of the 2024 ACM Conference on International Computing Education Research - Volume 1}} (Melbourne, VIC, Australia) \emph{(\bibinfo{series}{ICER '24})}. \bibinfo{publisher}{Association for Computing Machinery}, \bibinfo{address}{New York, NY, USA}, \bibinfo{pages}{469–486}.
\newblock
\showISBNx{9798400704758}
\urldef\tempurl%
\url{https://doi.org/10.1145/3632620.3671116}
\showDOI{\tempurl}


\bibitem[P\u{a}durean et~al\mbox{.}(2024)]%
        {padurean2024bugspotter}
\bibfield{author}{\bibinfo{person}{Victor-Alexandru P\u{a}durean}, \bibinfo{person}{Paul Denny}, {and} \bibinfo{person}{Adish Singla}.} \bibinfo{year}{2024}\natexlab{}.
\newblock \bibinfo{title}{BugSpotter: Automated Generation of Code Debugging Exercises}.
\newblock
\newblock
\showeprint[arxiv]{2411.14303}~[cs.SE]
\urldef\tempurl%
\url{https://arxiv.org/abs/2411.14303}
\showURL{%
\tempurl}


\bibitem[Sarsa et~al\mbox{.}(2022)]%
        {sarsa2022automatic}
\bibfield{author}{\bibinfo{person}{Sami Sarsa}, \bibinfo{person}{Paul Denny}, \bibinfo{person}{Arto Hellas}, {and} \bibinfo{person}{Juho Leinonen}.} \bibinfo{year}{2022}\natexlab{}.
\newblock \showarticletitle{Automatic Generation of Programming Exercises and Code Explanations Using Large Language Models}. In \bibinfo{booktitle}{\emph{Proceedings of the 2022 ACM Conference on International Computing Education Research - Volume 1}} (Lugano and Virtual Event, Switzerland) \emph{(\bibinfo{series}{ICER '22})}. \bibinfo{publisher}{Association for Computing Machinery}, \bibinfo{address}{New York, NY, USA}, \bibinfo{pages}{27–43}.
\newblock
\showISBNx{9781450391948}
\urldef\tempurl%
\url{https://doi.org/10.1145/3501385.3543957}
\showDOI{\tempurl}


\bibitem[Savelka et~al\mbox{.}(2023)]%
        {savelka2023can}
\bibfield{author}{\bibinfo{person}{Jaromir Savelka}, \bibinfo{person}{Arav Agarwal}, \bibinfo{person}{Christopher Bogart}, \bibinfo{person}{Yifan Song}, {and} \bibinfo{person}{Majd Sakr}.} \bibinfo{year}{2023}\natexlab{}.
\newblock \showarticletitle{Can generative pre-trained transformers (gpt) pass assessments in higher education programming courses?}. In \bibinfo{booktitle}{\emph{Proceedings of the 2023 Conference on Innovation and Technology in Computer Science Education V. 1}}. \bibinfo{pages}{117--123}.
\newblock


\bibitem[Sheese et~al\mbox{.}(2024)]%
        {sheese2024patterns}
\bibfield{author}{\bibinfo{person}{Brad Sheese}, \bibinfo{person}{Mark Liffiton}, \bibinfo{person}{Jaromir Savelka}, {and} \bibinfo{person}{Paul Denny}.} \bibinfo{year}{2024}\natexlab{}.
\newblock \showarticletitle{Patterns of Student Help-Seeking When Using a Large Language Model-Powered Programming Assistant}. In \bibinfo{booktitle}{\emph{Proceedings of the 26th Australasian Computing Education Conference}}. \bibinfo{pages}{49--57}.
\newblock


\bibitem[Tang et~al\mbox{.}(2024)]%
        {tang2024science}
\bibfield{author}{\bibinfo{person}{Ruixiang Tang}, \bibinfo{person}{Yu-Neng Chuang}, {and} \bibinfo{person}{Xia Hu}.} \bibinfo{year}{2024}\natexlab{}.
\newblock \showarticletitle{The Science of Detecting LLM-Generated Text}.
\newblock \bibinfo{journal}{\emph{Commun. ACM}} \bibinfo{volume}{67}, \bibinfo{number}{4} (\bibinfo{year}{2024}), \bibinfo{pages}{50--59}.
\newblock


\bibitem[Tran et~al\mbox{.}(2023)]%
        {10342898}
\bibfield{author}{\bibinfo{person}{Andrew Tran}, \bibinfo{person}{Kenneth Angelikas}, \bibinfo{person}{Egi Rama}, \bibinfo{person}{Chiku Okechukwu}, \bibinfo{person}{David~H. Smith}, {and} \bibinfo{person}{Stephen MacNeil}.} \bibinfo{year}{2023}\natexlab{}.
\newblock \showarticletitle{Generating Multiple Choice Questions for Computing Courses Using Large Language Models}. In \bibinfo{booktitle}{\emph{2023 IEEE Frontiers in Education Conference (FIE)}}. \bibinfo{pages}{1--8}.
\newblock
\urldef\tempurl%
\url{https://doi.org/10.1109/FIE58773.2023.10342898}
\showDOI{\tempurl}


\bibitem[Vadaparty et~al\mbox{.}(2024)]%
        {vadaparty2024cs1-llm}
\bibfield{author}{\bibinfo{person}{Annapurna Vadaparty}, \bibinfo{person}{Daniel Zingaro}, \bibinfo{person}{David~H. Smith~IV}, \bibinfo{person}{Mounika Padala}, \bibinfo{person}{Christine Alvarado}, \bibinfo{person}{Jamie Gorson~Benario}, {and} \bibinfo{person}{Leo Porter}.} \bibinfo{year}{2024}\natexlab{}.
\newblock \showarticletitle{CS1-LLM: Integrating LLMs into CS1 Instruction}. In \bibinfo{booktitle}{\emph{Proceedings of the 2024 on Innovation and Technology in Computer Science Education V. 1}} (Milan, Italy) \emph{(\bibinfo{series}{ITiCSE 2024})}. \bibinfo{publisher}{Association for Computing Machinery}, \bibinfo{address}{New York, NY, USA}, \bibinfo{pages}{297–303}.
\newblock
\showISBNx{9798400706004}
\urldef\tempurl%
\url{https://doi.org/10.1145/3649217.3653584}
\showDOI{\tempurl}


\bibitem[Wu et~al\mbox{.}(2023)]%
        {wu2023survey}
\bibfield{author}{\bibinfo{person}{Junchao Wu}, \bibinfo{person}{Shu Yang}, \bibinfo{person}{Runzhe Zhan}, \bibinfo{person}{Yulin Yuan}, \bibinfo{person}{Derek~F Wong}, {and} \bibinfo{person}{Lidia~S Chao}.} \bibinfo{year}{2023}\natexlab{}.
\newblock \showarticletitle{A survey on llm-gernerated text detection: Necessity, methods, and future directions}.
\newblock \bibinfo{journal}{\emph{arXiv preprint arXiv:2310.14724}} (\bibinfo{year}{2023}).
\newblock


\bibitem[W{\"u}nsche et~al\mbox{.}(2018)]%
        {wunsche2018automatic}
\bibfield{author}{\bibinfo{person}{Burkhard~C W{\"u}nsche}, \bibinfo{person}{Zhen Chen}, \bibinfo{person}{Lindsay Shaw}, \bibinfo{person}{Thomas Suselo}, \bibinfo{person}{Kai-Cheung Leung}, \bibinfo{person}{Davis Dimalen}, \bibinfo{person}{Wannes van~der Mark}, \bibinfo{person}{Andrew Luxton-Reilly}, {and} \bibinfo{person}{Richard Lobb}.} \bibinfo{year}{2018}\natexlab{}.
\newblock \showarticletitle{Automatic assessment of OpenGL computer graphics assignments}. In \bibinfo{booktitle}{\emph{Proceedings of the 23rd annual ACM conference on innovation and technology in computer science education}}. \bibinfo{pages}{81--86}.
\newblock


\bibitem[W{\"u}nsche et~al\mbox{.}(2019)]%
        {wunsche2019coderunnergl}
\bibfield{author}{\bibinfo{person}{Burkhard~C W{\"u}nsche}, \bibinfo{person}{Edward Huang}, \bibinfo{person}{Lindsay Shaw}, \bibinfo{person}{Thomas Suselo}, \bibinfo{person}{Kai-Cheung Leung}, \bibinfo{person}{Davis Dimalen}, \bibinfo{person}{Wannes van~der Mark}, \bibinfo{person}{Andrew Luxton-Reilly}, {and} \bibinfo{person}{Richard Lobb}.} \bibinfo{year}{2019}\natexlab{}.
\newblock \showarticletitle{CodeRunnerGL-An interactive web-based tool for computer graphics teaching and assessment}. In \bibinfo{booktitle}{\emph{2019 International Conference on Electronics, Information, and Communication (ICEIC)}}. IEEE, \bibinfo{pages}{1--7}.
\newblock


\end{thebibliography}
\end{document}